\documentclass[aps,twocolumn,preprintnumbers,floatfix,nofootinbib]{revtex4}

\usepackage{graphicx}

\usepackage{amsmath}
\usepackage{amsfonts}
\usepackage{amssymb}

\usepackage[utf8]{inputenc}

\usepackage{amsmath}
\usepackage{amsfonts}
\usepackage{amssymb}
\usepackage{color}

\def\be{\begin{equation}}
\def\ee{\end{equation}}
\def\bea{\begin{eqnarray}}
\def\eea{\end{eqnarray}}

\def\<{\langle}
\def\>{\rangle}

\def\mt{{\mathcal{T}}}
\def\LI{{\mathrm{LI}}}
\def\nlsm{nl$\sigma$m}

\def\mt{{\mathcal{T}}}
\def\LI{{\mathrm{LI}}}

\newcommand*\DAl{\mathop{}\!\mathbin\Box}

\def\slashchar#1{\setbox0=\hbox{$#1$}           
   \dimen0=\wd0                                 
   \setbox1=\hbox{/} \dimen1=\wd1               
   \ifdim\dimen0>\dimen1                        
      \rlap{\hbox to \dimen0{\hfil/\hfil}}      
      #1                                        
   \else                                        
      \rlap{\hbox to \dimen1{\hfil$#1$\hfil}}   
      /                                         
   \fi}

\begin{document}


\title{Ward Identity and Scattering Amplitudes for Nonlinear Sigma Models}

\author{Ian Low$^{\, a,b,c}$ and Zhewei Yin$^{\, b}$}
\affiliation{
\mbox{$^a$ High Energy Physics Division, Argonne National Laboratory, Argonne, IL 60439, USA}\\
\mbox{$^b$ Department of Physics and Astronomy, Northwestern University, Evanston, IL 60208, USA} \\
\mbox{$^c$ Theoretical Physics Department, CERN, 1211 Geneva 23, Switzerland}
}

\begin{abstract}
We present a  Ward identity for nonlinear sigma models using generalized nonlinear shift symmetries, without introducing current algebra or coset space. The Ward identity constrains  correlation functions of the sigma model  such that the Adler's zero is guaranteed for $S$-matrix elements, and gives rise to a subleading single soft theorem that is valid at the quantum level and to all orders in  the Goldstone decay constant.  For tree amplitudes, the Ward identity leads to a novel Berends-Giele recursion relation as well as an explicit form of the   subleading single soft factor. Furthermore, interactions of the cubic biadjoint scalar theory associated with the single soft limit, which was previously discovered using the Cachazo-He-Yuan representation of tree amplitudes, can be seen to emerge from matrix elements of  conserved currents  corresponding to the  generalized shift symmetry.
\end{abstract}


\maketitle

\section{Introduction}
\label{sec:introduction}
Nonlinear sigma models (\nlsm) \cite{GellMann:1960np} have  wide-ranging applications in many branches of physics. It was realized  early on that spontaneously broken symmetries play a central role in understanding the dynamics of \nlsm. Such a realization was  embodied in the current algebra approach \cite{Treiman:1986ep}, where currents corresponding to broken symmetry generators and their commutators allowed for computations of pion scattering amplitudes and resulted in the celebrated “Adler's zeros" in  the single emission of soft pions \cite{Adler:1964um}. Modern formulation of \nlsm\  is based on the coset space construction by Callan, Coleman, Wess and Zumino (CCWZ) \cite{Coleman:1969sm,Callan:1969sn}, where the Goldstone bosons parameterize the coset manifold $G/H$ with $G$ being the spontaneously broken symmetry in the UV and $H$ the unbroken group in the IR.

Lately there has been a resurgence of efforts in understanding the infrared structure of quantum field theories, in particular in gravity and gauge theories \cite{Strominger:2017zoo}. The classic soft theorems \cite{PhysRev.140.B516} were re-derived using asymptotic symmetries and the related Ward identities, while the soft massless particles are interpreted as Goldstone bosons residing at the future null infinity \cite{Strominger:2013lka,He:2014cra,Strominger:2013jfa,He:2014laa}. 

On the other hand, our understanding of the  Goldstone bosons in \nlsm\ had stayed at the same level as in the 1960's, until Ref.~\cite{ArkaniHamed:2008gz} studied the double soft emission of Goldstone bosons in the context of scattering amplitudes, which sparked new efforts in this direction \cite{Kampf:2013vha,Cheung:2014dqa,Cheung:2015ota}. More recently Ref.~\cite{Cachazo:2016njl} studied the subleading single soft limit of tree-level amplitudes in a variety of theories exhibiting Adler's zeros using the Cachazo-He-Yuan (CHY) representation of scattering equations \cite{Cachazo:2013hca,Cachazo:2013iea,Cachazo:2014xea}. They found in each case the subleading single soft factor can be interpreted as on-shell tree-amplitudes of a mysterious extended theory. Only the CHY representation of tree amplitudes in the extended theory is given, and little  is known regarding how the extended theory emerges. For \nlsm, the extended theory turns out to be a theory of cubic biadjoint scalars interacting with the Goldstone bosons.

In this work we aim to provide a common thread through the  different perspectives on the infrared dynamics of Goldstone bosons. We  first present a Ward identity governing the correlation functions of \nlsm\ such that the Adler's zero is guaranteed for $S$-matrix elements. The Ward identity is derived using nonlinear shift symmetries \cite{Low:2014nga,Low:2014oga}, which makes transparent the infrared universality of the  result, regardless of the underlying coset space. Using the identity we derive a single soft theorem, beyond the Adler's zero, that is valid both at the quantum level and to all orders in the Goldstone decay constant. For tree-level  amplitudes, we obtain a novel set of Berends-Giele recursion relations, which leads to the subleading single-soft factor of flavor-ordered tree amplitudes. In particular, the derivation shed lights on the emergence of the extended cubic biadjoint scalar theory uncovered using the CHY approach in Ref.~\cite{Cachazo:2016njl}.

\section{The Ward Identity}
\label{sec:Ward}
In Refs.~\cite{Low:2014nga,Low:2014oga} a new approach  to constructing the effective Lagrangian for \nlsm\ was proposed, without recourse to the current algebra or the coset construction. It is based on the simple observation that, for a spontaneously broken $U(1)$ symmetry, the effective Lagrangian for the sole Goldstone boson $\pi(x)$ can be constructed by imposing the shift symmetry:
\be
\label{eq:const}
\pi(x) \to \pi(x) + \epsilon \ ,
\ee
where $\epsilon$ is a constant. The constant shift symmetry enforces the  Adler's zero condition. For a non-trivial unbroken group $H$, there  are multiple Goldstone bosons $\pi^a(x)$ furnishing a linear representation of $H$ and the constant shift symmetry is enlarged to respect both the Adler's zero condition and the linearly realized $H$ symmetry.
Choosing a basis such that generators of $H$,  $(T^i)_{ab}$, are purely imaginary and anti-symmetric, and  adopting the bra-ket notation to define $|T^i\pi\rangle = T^i |\pi\rangle$, Eq.~(\ref{eq:const}) can be generalized to \cite{Low:2014nga,Low:2014oga}:
\be
\label{eq:Tdef}
|\pi\rangle \to |\pi\rangle  + \sum_{k=0}^\infty {a_k} \mt^k |\epsilon\rangle \ , \quad {\cal T} \equiv \frac1{f^2} |T^i \pi\rangle\langle \pi T^i |\ ,
\ee
where $a_k$ are numerical constants, $|\epsilon\rangle$ is a constant vector, and $f$ is the Goldstone decay constant.  By imposing the Adler's zero condition, an effective Lagrangian for the Goldstone bosons can be constructed, without specifying the broken group $G$ in the UV, up to the overall normalization of  $f$. The construction makes it clear that interactions of Goldstone bosons are universal in the IR and insensitive to the coset structure $G/H$. The highly nonlinear nature of the Goldstone interactions only serves two purposes: 1) fulfilling the Adler's zero condition and 2) linearly realizing the unbroken group $H$. The leading two-derivative Lagrangian is \cite{Low:2014nga,Low:2014oga}
\be
{\cal L}^{(2)} = \frac12 \langle {\cal D}_\mu \pi | {\cal D}^\mu \pi\rangle \ ,\quad |{\cal D}_\mu \pi\rangle = \frac{\sin\sqrt{\cal T}}{\sqrt{\cal T}} |\partial_\mu \pi\rangle \ .
\ee
Using the universality in  Goldstone interactions, it is possible to derive the generalized nonlinear shift in Eq.~(\ref{eq:Tdef}) \cite{forthcoming}:
\be
\label{eq:NLshift}
|\pi\rangle \to |\pi\rangle+{F}_1(\mt) |\epsilon\rangle \ , \quad  {F}_1(\mt)= \sqrt{\cal T} \cot \sqrt{\cal T}\ ,
\ee
under which  the \nlsm\ Lagrangian is invariant.

It is now straightforward to derive the Ward identity corresponding Eq.~(\ref{eq:NLshift}) in path integral, by promoting the global transformation into a local one \cite{Peskin:1995ev}: $|\epsilon\rangle \to |\epsilon(x)\rangle$, which  amounts to a change of variable in evaluating the path integral and leads to the  Ward identity:
\bea
\label{eqwig} 
&&\!\!\!\!\!\!\!\!\!\!\!\!\! i\, \partial_\mu \langle 0| \left\{  \left[F_2(\mt) \right]_{ab}\partial^\mu \pi^b \right\} (x) \prod_{i=1}^n \pi^{a_i} (x_i)|0 \rangle= \nonumber \\
&&\!\!\!\!\!\!\!\!\!\!\!\!\!\! \sum_{r=1}^n \Delta_r \langle 0| \pi^{a_1} (x_1) \cdots \left\{ \left[F_1(\mt)\right](x_r) \right\}_{a_r a}\cdots \pi^{a_n} (x_n)|0 \rangle ,
\eea
where $\partial_\mu=\partial/\partial x^\mu$ and
\be
\Delta_r = \delta^{(4)}(x-x_r) \ ,\quad
F_2(\mt) = \frac{\sin \sqrt{\mt} \cos \sqrt{\mt}}{\sqrt{\mt}}.
\ee
Since we have not invoked any specific coset structure, Eq.~(\ref{eqwig}) is universal. It is worth reiterating that we have only invoked the Adler's zero condition and the linearly realized unbroken symmetry $H$. This is in contrast with the  usual vector and axial Ward identities considered in current algebra, which  assumes the existence of  broken symmetry generators as well as the associated current commutators.

\section{A Berends-Giele  Relation}
\label{sec:BGrecursion}

The semi-on-shell amplitude is defined as
\bea
\label{eq:semidef}
J^{a_1 \cdots a_n, a} (p_1, \cdots ,p_n) = \langle 0| \pi^a (0) | \pi^{a_1}(p_1) \cdots \pi^{a_n}(p_n)\rangle . 
\eea
Such objects were considered first by Berends and Giele in Ref.~\cite{Berends:1987me} as building blocks for computing $S$-matrix elements in Yang-Mills theories. In $SU(N)$ \nlsm\ they were studied in  Ref.~\cite{Kampf:2013vha} and a Berends-Giele type recursion relation was proposed using Feynman vertices from an effective Lagrangian.

Eq.~(\ref{eq:semidef}) can be obtained from a $(n+1)$-point correlation function via the Lehmann-Symanzik-Zimmermann (LSZ) reduction on $n$ of the Goldstone fields \cite{Peskin:1995ev}. We define 
\bea
\LI \equiv \left(\frac{i}{\sqrt{Z}}\right)^{n} \int d^4 x\, e^{-i q \cdot x}  \prod_{i=1}^n \int d^4 x_i\, e^{-i p_i \cdot x_i}\, \DAl_i \ ,
\eea
and  perform the LSZ reduction on the $n$ Goldstone bosons  by taking the on-shell limit $p_i^2 \to 0$, $i=1,\cdots, n$. We have also performed the Fourier transform with respect to $q$ in the above, so that $q = - \sum_{i=1}^n p_i$ after the integration.  In doing so,  observe that the right-hand side (RHS) of Eq.~(\ref{eqwig})  contains only $(n-1)$ single particle poles  and, therefore, vanishes. The left-hand side (LHS) can be expanded in a power series in $1/f^2$ using $F_2(x) =\sum_k(-4)^k x^k/(2k+1)!$, the first of which is exactly 
\bea
&& \lim_{p_1^2 \to 0} \cdots \lim_{p_n^2 \to 0}\ \LI\ \partial_\mu \<0|\partial^\mu \pi^a (x) \prod_{i=1}^n \pi^{a_i} (x_i) |0\> \nonumber \\
& &\qquad = -q^2 J^{a_1 \cdots a_n, a} (p_1, \cdots ,p_n) \ ,
\eea
while the higher order terms are matrix elements of the form $\<0| \tilde{O}^a_k(q) | \pi^{a_1}(p_1) \cdots \pi^{a_n}(p_n)\>$, where
\be
\label{eq:Okdef}
\tilde{O}^a_k(q) =  \int d^4 x\, e^{-i q \cdot x} \partial_\mu\left\{ \left[{\cal T}^k(x)\right]_{ab} \partial^\mu \pi^b(x)\right\}  \ .
\ee
The Ward identity now turns into
\bea
\label{eq:quantward}
 &&\!\!\! q^2 J^{a_1 \cdots a_n, a} (p_1, \cdots ,p_n) \nonumber\\
&&\!\!\!\!\!\!\!\!\!= \sum_{k=1}^{\infty} \frac{(-4)^k}{(2k+1)!} \<0| \tilde{O}^a_k(q) | \pi^{a_1}(p_1) \cdots \pi^{a_n}(p_n)\> \ ,
\eea
which is valid at the quantum level. Since $\tilde{O}^a_k(q)$ is  proportional to $q_\mu$,  the Adler's zero is manifest  when $q_\mu \to 0$.

At the classical level, Eq.~(\ref{eq:quantward}) can be turned into a recursion relation among tree-level semi-on-shell amplitudes. Let's define a  $(2k+1)$-point tree vertex  from  $\tilde{{\cal O}}_k^a$,
\bea
\label{eq:vertex}
&&V^{a_1 \cdots a_{2k+1},a} (p_1, \cdots, p_{2k+1}) \nonumber\\
&&= \frac{-i(-4)^{k} }{(2k+1) ! f^{2k}}  \sum_\sigma C^{a_1 \cdots a_{2k+1}, a}_\sigma\ q \cdot p_{\sigma (2k+1)}\ ,
\eea
where $\sigma$ is a permutation of $\{ 1,2, \cdots, 2k+1 \}$ and
\bea
\label{eq:colordef}
&& C^{a_1 \cdots a_{2k+1}, a}_\sigma \equiv  T^{i_1}_{a a_{\sigma (1)}} T^{i_1}_{a_{\sigma (2)} b_1} T^{i_2}_{b_1 a_{\sigma (3)}} T^{i_2}_{a_{\sigma (4)} b_2} \cdots\nonumber\\
&&\qquad \times \ T^{i_k}_{b_{n-1} a_{\sigma (2k-1)}} T^{i_k}_{a_{\sigma (2k)} a_{\sigma (2k+1)}} \ .
\eea
 Then the Berends-Giele recursion relation is obtained from Eq.~(\ref{eq:quantward}) by connecting the vertex in Eq.~(\ref{eq:vertex}) with either an external leg or the off-shell leg of a sub-semi-on-shell amplitude, whose on-shell legs are a subset of $\{p_1,\cdots, p_n\}$. In the end we arrive at 
 \bea
 \label{eq:bgrelation}
 && q^2 J^{a_1 \cdots a_{n}, a} (p_1, \cdots ,p_{n}) = \nonumber\\
 &&\quad i \sum_{k=1}^{[n/2]} \sum_{\{d^\ell\}} V^{b_1\cdots b_{2k+1},a}(q_{d^{\ell,1}},\cdots,q_{d^{\ell,2k+1}})\nonumber\\
 &&\qquad \times \prod_{i=1}^{2k+1} J^{a_{d^{\ell,i}_1} a_{d^{\ell,i}_2}\cdots, b_i}(p_{d^{\ell,i}_1}, p_{d^{\ell,i}_2},\cdots)\ .
 \eea
Here $d^\ell=\{d^{\ell,i}\}$ is a way to divide $\{1,2,\cdots, n\}$ into $2k+1$ disjoint, non-ordered subsets.
The $j$th element of $d^{\ell,i}$ is denoted by $d^{\ell,i}_j$ and $q_{d^{\ell,i}}=\sum_j p_{d^{\ell,i}_j}$. 
 
 Semi-on-shell amplitudes in \nlsm\ are not invariant under field redefinitions and depend on the particular parameterization employed to write down the  Lagrangian. It is then interesting to highlight the difference between  Eq.~(\ref{eq:bgrelation}) from that obtained using Feynman diagrams in Ref.~\cite{Kampf:2013vha}. The vertex in Eq.~(\ref{eq:bgrelation}) arises from the operator insertion of $\tilde{\cal O}^a_k(q)$, which carries momentum injection of $q^\mu$.  In fact, as we will see, the cubic interaction of the extended biadjoint scalar is given by the matrix element of $\tilde{\cal O}^a_1(0)$, which is a three-point vertex.  In addition, the explicit Adler's zero in the limit $q^\mu\to 0$ in Eq.~(\ref{eq:vertex})  greatly facilitates the calculation of the subleading single soft limit of the \nlsm, which we turn to next.

\section{The Subleading Single Soft Limit}

On-shell  amplitudes are further obtained from the semi-on-shell amplitudes by 
\be
M^{a_1\cdots a_{n+1}}= \lim_{q^2\to 0} -\frac{1}{\sqrt{Z}} q^2 J^{a_1\cdots a_n, a_{n+1}}\ ,
\ee
where $q = -(p_1+\cdots+p_{n})\equiv p_{n+1}$ is the momentum of the $(n+1)$-th leg. Using Eq.~(\ref{eq:quantward}) it is simple to derive the single soft limit  in $p_{n+1}\to \tau p_{n+1}$, $\tau\to 0$:
\bea
\label{eq:quantsoft}
 &&\!\!\! M^{a_1 \cdots a_{n+1}} \to \frac{1}{\sqrt{Z}} \sum_{k=1}^{\infty} \frac{-(-4)^{k}}{(2k+1)!} \nonumber\\
&&\!\!\!\!\!\!\!\!\!\!\!\!\!\! \times\ \tau\, \<0| \int d^4x\, [{\cal T}^k(x)]_{ab}\ ip_{n+1}\cdot \partial\, \pi^b(x) | \pi^{a_1} \cdots \pi^{a_n}\>,
\eea
This is the subleading single soft theorem in \nlsm, valid at the quantum level. Notice there is no momentum injection at this order in $\tau$ and the operator behaves just like a ``normal" Feynman vertex, which hints at interpreting the matrix element  as scattering amplitudes.

Ref.~\cite{Cachazo:2016njl} studied the flavor-ordered tree amplitudes, which we now proceed to consider. 
Recall that we chose the generator $T^i$ of the unbroken group $H$ to be purely imaginary and anti-symmetric. The equivalence of our approach with the coset space construction, which  introduces broken symmetry generators $X^a$, is readily established upon the  identification \cite{Low:2014nga,Low:2014oga}
\be
(T^i)_{ab}  =- i f^{iab}\ ,
\ee
where $[T^i,X^a]=if^{iab}X^b$ and $[X^a,X^b]=if^{abi}T^i$ for symmetric cosets. Using the normalization ${\rm Tr}(X^aX^b)=\delta^{ab}$ one can  show that the color factor in Eq.~(\ref{eq:colordef}) becomes
\bea
&&C^{a_1 \cdots a_{2k+1}, a}_\sigma = \nonumber \\
&&\!\!\!\!\!\!\!\!\!\!\!\!\!\!\!\!\!\!{\rm Tr} ( [[\cdots [ [X^a, X^{a_{\sigma (1)}}], X^{a_{\sigma (2)}}], \cdots ], X^{a_{\sigma (2k)}}] X^{a_{\sigma (2k+1)}} ).
\eea
A flavor-ordered vertex $V(1,2,\cdots,2k+1)$ from Eq.~(\ref{eq:vertex}) can now be defined
\bea
&&V^{a_1 \cdots a_{2k+1},a} (p_1, \cdots, p_{2k+1}) \equiv\nonumber\\
 &&\!\!\!\!\!\!\!\!\!\!\!\!\!\!\!\!\! \sum_{\sigma} {\rm Tr} ( X^a X^{a_{\sigma (1)}} \cdots X^{a_{\sigma (2k+1)}} ) V_\sigma(p_1,\cdots,p_{2k+1})\ .
\eea
Furthermore, using the notation $V(1,\cdots, 2k+1)= V_\sigma(p_1,\cdots,p_{2k+1})$ for $\sigma=$ identity,
\bea
&&V(1,2,\cdots, 2k+1) =\nonumber\\
&&\qquad \frac{-i(-4)^k}{(2k+1)!f^{2k}}\sum_{j=0}^{2k} \left(\begin{array}{c}
2k\\
j
\end{array} \right) (-1)^{j} q \cdot p_{j+1}\ ,
\label{eqovco} 
\eea
where   $q$ is the momentum injection at the vertex.

Define the flavor-ordered semi-on-shell amplitude $J_\sigma(p_1,\cdots,p_{n})$ and $J(1,\cdots,n)$ similarly, Eq.~(\ref{eq:bgrelation}) gives
\bea
\label{eq:bgco}
&&q^2 J(1,2,\cdots,n)= i \sum_{k=1}^{[n/2]}  \sum_{\{l_m \}} V(q_{l_{1}},\cdots,q_{l_{2k+1}})\nonumber\\
&&\qquad \times  \prod_{m=1}^{2k+1} J (l_{m-1}+1, \cdots , l_m)\ ,
 \eea
where $l_m$ is a  splitting of the ordered set $\{1,2,\cdots, n\}$ into $2k+1$ non-empty ordered subsets $\{l_{m-1}+1,l_{m-1}+2,\cdots,l_m\}$  (here $l_0 = 1$ and $l_{2k+1} = n$). Moreover, $q_{l_{m}} = \sum_{i=l_{m-1}+1}^{l_m} p_i$. Eq.~(\ref{eq:bgco}) has a clear diagrammatic interpretation: $J(1,2,\cdots,n)$ is consisted of subamplitudes connecting to  $V(q_{l_{1}},\cdots,q_{l_{2k+1}})$.

The LHS of Eq.~(\ref{eq:bgco}) can be turned  into an on-shell amplitude by taking $q^\mu=-\sum_{i=1}^n p_i^\mu$ on-shell. Together with momentum conservation, the flavor-ordered vertex in Eq.~(\ref{eqovco}) can be written as
\bea
&&\!\!\!\!\!\!V (1,2,\cdots, 2k+1) =  \nonumber\\
&& \!\!\!\!\!\!\!\!\frac{-i(-4)^k}{(2k+1)!f^{2k}} \sum_{j=1}^{2k-1}\left[ \left(\begin{array}{c}
2k\\
j
\end{array} \right) (-1)^{j} -1 \right] q\cdot p_{j+1}\ ,
\eea
and  Eq.~(\ref{eq:bgco}) becomes
\bea
&&M(1,2,\cdots, n+1)= \sum_{k=1}^{[n/2]}  \frac{-(-4)^k}{(2k+1)!f^{2k}}\nonumber\\
&&\qquad \times \sum_{\{l_m \}}\sum_{j=1}^{2k-1} \left[ \left(\begin{array}{c}
2k\\
j
\end{array} \right) (-1)^{j} -1 \right] p_{n+1} \cdot q_{l_{j+1}}\nonumber\\
&&\qquad \times \prod_{m=1}^{2k+1} J (l_{m-1}+1, \cdots , l_m)\ ,
\label{eqst2}
\eea
where 
$p_{n+1}=q=-\sum_{i=1}^n p_i$.

At this stage Eq.~(\ref{eqst2}) is exact, having only taken the on-shell limit $p_{n+1}^2=0$. If we further take the soft limit, $p_{n+1}\to \tau p_{n+1}$, $\tau \to 0$, the RHS of Eq.~(\ref{eqst2}) starts at linear order in $\tau$, in accordance with the Adler's zero condition. Notice that at ${\cal O}(\tau)$, one can simply drop the $\tau$ dependence in the subamplitudes, by requiring  $\sum_{i=1}^n p_i = 0$. This is the next-to-leading order single soft factor  of flavor-ordered tree amplitudes in \nlsm.

\section{The CHY Interpretation}

In Ref.~\cite{Cachazo:2016njl}  the subleading single soft limit of flavor-ordered tree amplitudes in \nlsm\  is studied using the CHY formulation of scattering equations \cite{Cachazo:2013hca,Cachazo:2013iea,Cachazo:2014xea}. The single soft limit is interpreted as relating the $(n+1)$-point  amplitudes in \nlsm\ to the $n$-point amplitudes of a related, but different theory containing cubic interactions of biadjoint scalars. Specifically, at  ${\cal O}(\tau)$,  the proposal is
\be
\label{eq:chy}
M(\mathbb{I}_{n+1}) = \tau \sum_{i=2}^{n-1} s_{n+1,i}\ M^{\text{\nlsm}\oplus \phi^3}(\mathbb{I}_{n}|1,n,i)\ ,
\ee 
where $s_{ij}=2p_i\cdot p_j$.  $M(\mathbb{I}_{n+1})$ is the $(n+1)$-point flavor-ordered amplitude in \nlsm\ with the ordering $\mathbb{I}_{n+1}=\{1,2,\cdots, n+1\}$ and $M^{\text{\nlsm}\oplus \phi^3}(\mathbb{I}_{n}|1,n, i)$ denotes the $n$-point amplitudes of  \nlsm\ interacting with a cubic biadjoint scalar, where $\{1,n,i\}$ is the flavor-ordering of the second adjoint index in the biadjoint scalar.  Little is known about the nature of this ``extended theory," and only the CHY representation of the flavor-ordered on-shell amplitudes is given.

Our results in the previous sections shed light on the interactions, in particular the Feynman vertices, of the extended theory. First of all, the emergence of a cubic scalar interaction is evident already in Eq.~(\ref{eq:quantsoft}). Using the 4-point amplitude as an example and set $p_4=q$ as the soft momentum, the full tree amplitude from Eq.~(\ref{eq:quantsoft}) is 
\bea
\label{eq:4ptsoft}
 && M^{a_1 a_2 a_3 a_4} \nonumber\\
 &=&\!\!\!\tau\, \frac{2}{3f^2} (T^i)_{a_4 r}(T^i)_{sb}  \<0| \int d^4x\,\pi^r\pi^s\, iq\cdot \partial\, \pi^b | \pi^{a_1} \pi^{a_2} \pi^{a_3} \> \nonumber\\
 &=&\!\!\! \tau\,\frac1{f^2} \sum_{\sigma} {\rm Tr}(X^{a_4}X^{a_{\sigma(1)}}X^{a_{\sigma(2)}}X^{a_{\sigma(3)}})  s_{4,\sigma(2)} ,
\eea
where we have used the on-shell condition, $q^2=0$, and momentum conservation, $q=-(p_1+p_2+p_3)$. Setting the decay constant $f = 1$, and extracting the  flavor-ordered single-soft factor  using the CHY proposal in Eq.~(\ref{eq:chy}), we obtain the cubic interaction
\be
\label{eq:3ptscalar}
M^{\text{\nlsm}\oplus \phi^3}(123|132) = -1 \ ,
\ee
which agrees with the CHY representation of 3-point amplitude given in Ref.~\cite{Cachazo:2016njl}. Tracing back the appearance of the cubic  interaction we see it is rooted in  the order $1/f^2$ term  in Eq.~(\ref{eqwig}), which is  a cubic operator.

To study Eq.~(\ref{eqst2}) in the context of the CHY proposal, which only gives the flavor-ordered tree amplitudes but not Feynman rules, we make the following observations regarding the flavor-ordered Feynman rule of the biadjoint scalar $\phi$:
1) no vertices exist with only one $\phi$, 2) a flavor-ordered $m$-point vertex containing  two $\phi$'s  has the same  flavor-ordered Feynman rule  as in the  \nlsm, and 3) a $(2k+1)$-point vertex involving three $\phi$'s has the following flavor-ordered Feynman rule 
\begin{align}
&V^{\text{\nlsm} \oplus \phi^3} (1, 2, \cdots ,j,\cdots,2k+1 | 1,2k+1,j) \nonumber\\
&\ \  =\frac{1}{2} \frac{-i(-4)^k}{(2k+1)!} \left[ \left(\begin{array}{c}
2k\\
j-1
\end{array} \right) (-1)^{j-1} -1 \right],
\end{align}
where $p_1$, $p_j$ and $p_{2k+1}$ are the momenta of $\phi$'s. Similar to the 3-point vertex in Eq.~(\ref{eq:3ptscalar}), the $(2k+1)$-point vertex can be seen as emerging from the order $1/f^{2k}$ term in the Ward identity in Eq.~(\ref{eqwig}).

Using these Feynman rules, it is possible to show that the coefficients of $s_{n+1,i}$ in Eq.~(\ref{eqst2}) are precisely the amplitudes in \nlsm$\oplus\phi^3$, in accordance with the CHY proposal in Eq.~(\ref{eq:chy}). In other words, these coefficients have consistent factorization and can be interpreted as scattering amplitudes \cite{forthcoming}.

\section{Conclusion}

In this work we have explored consequences of nonlinear shift symmetries in \nlsm\ and presented the associated Ward identity, which allowed us to study various aspects  of scattering amplitudes in \nlsm. In particular, we derived a next-to-leading order single soft theorem and studied the subleading single soft factor for flavor-ordered tree amplitudes, which provided a new perspective on the mysterious extended theory of cubic biadjoint scalars interacting with the Goldstone bosons.

There are many future directions. 
One example is whether the interpretation of an extended theory can be applied to the full scattering amplitudes of \nlsm, instead of just the flavor-ordered amplitudes. Naively there is an obstacle in doing so, since the LHS of Eq.~(\ref{eq:chy}) carries one flavor index, while the biadjoint amplitude in the RHS carries two flavor indices. Another possibility is to extend the Ward identity to shift symmetries involving spacetime, and understand their soft theorems and the associated extended theories. Additionally, there is a new formulation of \nlsm\ which makes the flavor-kinematic duality transparent \cite{Cheung:2016prv}, in which the subleading soft theorems and the cubic biadjoint scalars can be accommodated. It would be interesting to understand the connection with the shift symmetry perspective.

\begin{acknowledgments}
We would like to thank Anirudh Krovi for useful discussions and collaborations at the early stage, as well as Song He for reading and commenting on the manuscript. Z.Y. also acknowledges helpful discussions with Nima Arkani-Hamed. This work is supported in part by the U.S. Department of Energy under contracts No. DE-AC02-06CH11357 and No. DE-SC0010143.  
\end{acknowledgments}




\bibliography{references_amp}



\end{document}